\documentclass[twoside,twocolumn,9pt]{article}
\usepackage{extsizes}
\usepackage[super,sort&compress,comma]{natbib} 
\usepackage[version=3]{mhchem}
\usepackage[left=1.5cm, right=1.5cm, top=1.785cm, bottom=2.0cm]{geometry}
\usepackage{balance}
\usepackage{mathptmx}
\usepackage{sectsty}
\usepackage{graphicx} 
\usepackage{lastpage}
\usepackage[format=plain,justification=justified,singlelinecheck=false,font={stretch=1.125,small,sf},labelfont=bf,labelsep=space]{caption}
\usepackage{float}
\usepackage{fancyhdr}
\usepackage{fnpos}
\usepackage[english]{babel}
\addto{\captionsenglish}{%
  
}
\usepackage{array}
\usepackage{droidsans}
\usepackage{charter}
\usepackage[T1]{fontenc}
\usepackage[usenames,dvipsnames]{xcolor}
\usepackage{setspace}
\usepackage[compact]{titlesec}
\usepackage{hyperref}
\usepackage{enumerate} 
\usepackage{xcolor}
\usepackage{amssymb}

\definecolor{black}{RGB}{000,000,000}

\begin{document}

\pagestyle{fancy}
\thispagestyle{plain}
\fancypagestyle{plain}{

\renewcommand{\headrulewidth}{0pt}
}

\makeFNbottom
\makeatletter
\renewcommand\LARGE{\@setfontsize\LARGE{15pt}{17}}
\renewcommand\Large{\@setfontsize\Large{12pt}{14}}
\renewcommand\large{\@setfontsize\large{10pt}{12}}
\renewcommand\footnotesize{\@setfontsize\footnotesize{7pt}{10}}
\makeatother

\renewcommand{\thefootnote}{\fnsymbol{footnote}}
\renewcommand\footnoterule{\vspace*{1pt}%
\color{black}\hrule width 3.5in height 0.4pt \color{black}\vspace*{5pt}} 
\setcounter{secnumdepth}{5}

\makeatletter 
\renewcommand\@biblabel[1]{#1}            
\renewcommand\@makefntext[1]%
{\noindent\makebox[0pt][r]{\@thefnmark\,}#1}
\makeatother 
\renewcommand{\figurename}{\small{Fig.}~}
\sectionfont{\sffamily\Large}
\subsectionfont{\normalsize}
\subsubsectionfont{\bf}
\setstretch{1.125} 
\setlength{\skip\footins}{0.8cm}
\setlength{\footnotesep}{0.25cm}
\setlength{\jot}{10pt}
\titlespacing*{\section}{0pt}{4pt}{4pt}
\titlespacing*{\subsection}{0pt}{15pt}{1pt}

\fancyhead{}
\renewcommand{\headrulewidth}{0pt} 
\renewcommand{\footrulewidth}{0pt}
\setlength{\arrayrulewidth}{1pt}
\setlength{\columnsep}{6.5mm}
\setlength\bibsep{1pt}

\makeatletter 
\newlength{\figrulesep} 
\setlength{\figrulesep}{0.5\textfloatsep} 

\newcommand{\topfigrule}{\vspace*{-1pt}%
\noindent{\color{black}\rule[-\figrulesep]{\columnwidth}{1.5pt}} }

\newcommand{\botfigrule}{\vspace*{-2pt}%
\noindent{\color{black}\rule[\figrulesep]{\columnwidth}{1.5pt}} }

\newcommand{\dblfigrule}{\vspace*{-1pt}%
\noindent{\color{black}\rule[-\figrulesep]{\textwidth}{1.5pt}} }

\makeatother

\noindent\LARGE{\textbf{Bubble growth in a confined heated polymer: the example of safety glass}}

\vspace{0.3cm}  
\vspace{0.3cm}

\noindent\large{C. Arauz-Moreno,$^{\ast}$\textit{$^{a,b}$} K. Piroird,\textit{$^{b}$} and E. Lorenceau\textit{$^{a}$}} \\

 \noindent\normalsize{Laminated safety glass (LSG) is a composite assembly of glass and polyvinyl butyral (PVB), a viscoelastic polymer. LSG can be found in building facades and in all major forms of transportation. Yet, the assembly suffers from unwanted bubbles which are anathema to one of the most important features of glass: optical transparency. In here, we present an in-depth study of the reasons behind these bubbles, either during high-temperature quality control tests or normal glass operating conditions. We provide a physical model for bubble growth that deals with two gases, thermal effects on gas solubility and diffusivity, and a time-temperature dependent rheology. The model can be extended to n-component bubbles and various materials. By combining experiments and theory, we show that two gases are at play: air trapped in interfacial bubbles during lamination and water initially dissolved in the polymer bulk. Both gases work in tandem to induce bubble growth in finished assemblies of LSG provided that (i) the original bubble nucleus is large enough and (ii) the polymer relaxes (softens) sufficiently enough, especially at elevated temperatures. The latter constraints are relaxed in a condition we termed \textit{anomalous air oversaturation} that may even trigger a catastrophic, yet beautiful instability reminiscent of snowflakes or window frost.} \\

\renewcommand*\rmdefault{bch}\normalfont\upshape
\rmfamily
\section*{}
\vspace{-1cm}

\footnotetext{\textit{$^{a}$~Université Grenoble Alpes, CNRS, LIPhy, F-38000 Grenoble, France; E-mail: c.arauz\_moreno@icloud.com}}
\footnotetext{\textit{$^{b}$~Saint-Gobain Research Paris, F-93360 Aubervilliers, France. }}

\section{\label{sec:level1}Introduction}

Laminated safety glass (LSG) was invented by French chemist and artist, Edouard Benedictus (1878-1930) in what can only be described as a serendipitous discovery. As the lore suggests, in 1903 Benedictus knocked over a glass flask while on a ladder and much to his surprise, the flask did not break into smithereens after hitting the ground. Upon inspecting what surely was a curious occurrence at the time, he realized that a thin cellulose film held the glass together in place~\citep{nascimento2016first, panati2016panati, carrot2016polyvinyl}. Six years later, Benedictus would patent \textit{Triplex Glass}: two layers of glass bonded by a celluloid interlayer inspired by the broken flask of years prior. In his 1909 patent (FR405881), Benedictus identified the main advantages of his invention: superior impact resistance in comparison to monolithic glass, and when broken, the glass remains bonded to the interlayer. Despite its humble and lucky beginnings, laminated safety glass, now bonded by polyvinyl butyral (PVB), is currently a staple of modern life, from the world famous pyramid in the Louvre museum to Airbus's cockpit glass, passing by buildings, cars, bridges, balconies, and balustrades. 

It is not surprising then, that safety glass is the subject of active scientific research in terms of blast, bonding, delamination, and breakage performance~\citep{del2016determining,samieian2019bonding, fourton2020adhesion, chen2022pummel, muller2024imperfections, zhou2024post}, and that the PVB polymer has been extensively studied in terms of chemical composition, transport properties, deformation mechanisms, and mechanical response to name a few~\citep{corroyer2013characterization, arauz2023water,pauli2024simplified,yang2024deformation, lopez2019mechanical,centelles2021viscoelastic, arauz2022extended, stevels2020determination}. Yet, a pressing issue remains largely scientifically unexplored, one the industry recognizes and actively tests against: \textit{bubbles}. The latter may appear after manufacturing, during quality control tests, or simply over the glass lifetime. 

\begin{figure*}[]
\centering{\includegraphics[scale=0.6]{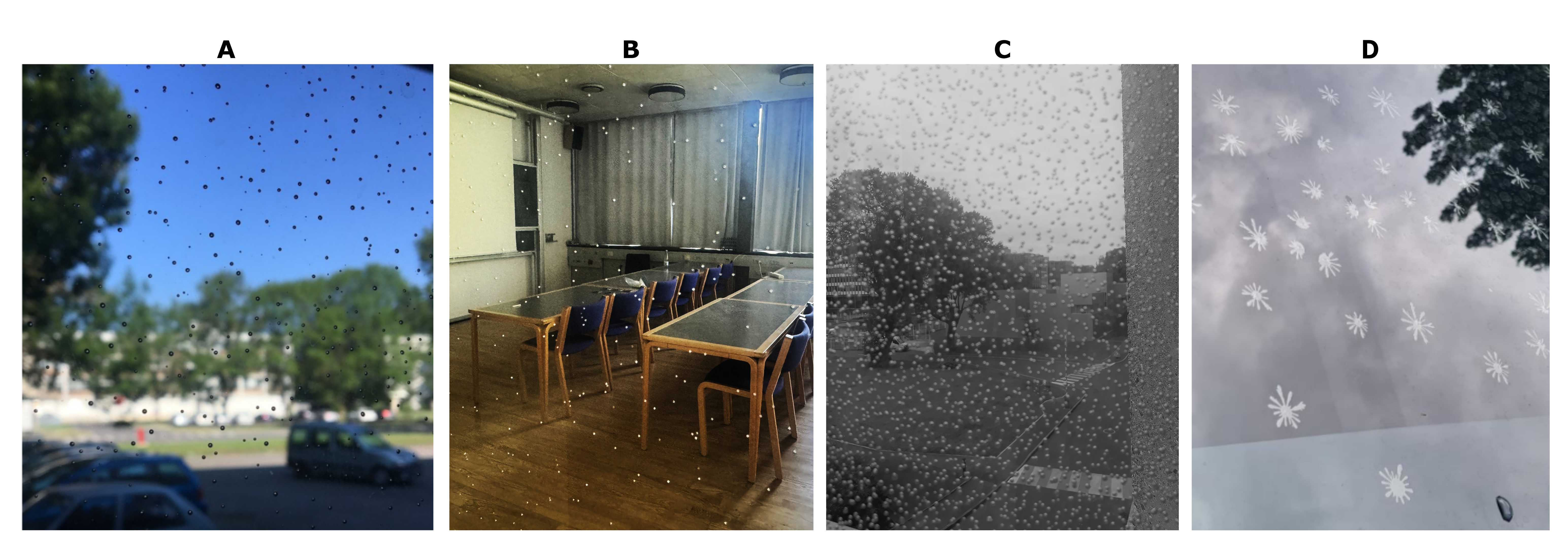}}
\caption{Photographs of bubbles in laminated safety glass at \textbf{A} the LIPhy laboratory in Grenoble (courtesy of C. Schune, Saint-Gobain), \textbf{B} Technical University of Denmark (DTU) (C. Arauz-Moreno), \textbf{C} Jussieu Campus, Sorbonne University in Paris (E. Lorenceau), and \textbf{D} window frost failure in building façade (K. Piroird). See $\S$~\ref{sec:phys_desc}-\ref{sec:air_main} for a thorough discussion on the physical mechanism behind these bubbles.} 
\label{fig1}
\end{figure*}

Bubbles can defeat the very purpose of glass which is optical transparency. To this end, \hyperref[fig1]{figure~1} presents a plurality of cases of safety glass failure by bubbling. The presence of such bubbles leads to entire production batches being scrapped during manufacturing, or worse, recalled from a customer site\textemdash a situation that leads to eventual losses, monetary or otherwise, for both suppliers and consumers. This situation comes with environmental implications because safety glass is difficult to recycle, both technically and economically.  

The likelihood of bubble formation is tested destructively using a so-called bake test (EN ISO 12543-4) wherein a piece of finished glass is placed inside an oven at 100°C for 16hrs to induce bubble formation. Presumably, the test indicates whether air\textemdash the presupposed culprit behind bubbles\textemdash is present in large quantities in the glass assembly. 

Yet, the nucleation and growth of bubbles in safety glass remains to a great extent unexplained despite the inherent multi-physics of the problem and the many interesting questions that surround them: how are bubbles nucleated? Is air the real (or only) culprit? What role do the flow properties of the viscoelastic PVB polymer play? Is the bake test an accurate quality control test? Can it be modified or improved? In this paper, we answer many of these questions by combining experiments with theory. In $\S$~\ref{sec:sci_dissect}, we provide a detailed scientific account of the lamination process using our purposefully made laboratory scale protocol. In doing so, we derive observations not readily accessible at the industrial scale in terms of glass optical appearance, mass transfer, adhesion, and polymer rheology. $\S$~\ref{sec:toy_exp} is an exemplary toy experiment, first of its kind to the best of our knowledge in the available literature, into the bubbling behaviour of a glass assembly during lamination. In $\S$~\ref{sec:phys_desc}, we present a physical model for the growth of bubbles in safety glass. The model, while being motivated by the problem at hand, is general enough in its treatment of mass transport, thermal effects, and rheological phenomena and can easily account for n-component bubbles, isothermal/non isothermal conditions, and even different mediums beyond PVB. In the remaining sections, to wit, $\S$~\ref{sec:nucleation},~\ref{sec:bake_test},~\ref{sec:water_partly},~\ref{sec:air_main} we deal with the subjects of bubble nucleation as well as the role of gases dissolved in the PVB bulk for bubble formation. 

\section{Scientifically Dissecting Glass Lamination}\label{sec:sci_dissect}

Laminated safety glass is constituted by two layers of float glass that are bonded together using polyvinyl butyral (PVB), a random chain, amorphous polymer of vinyl acetate, vinyl alcohol, and vinyl butyral that is commonly plasticized with triethylene glycol di(2-ethylhexanoate) to lower the glass transition temperature~\citep{elziere2019supramolecular,stevels2024ten}. PVB comes in different presentations. For example, Eastman produces rubbery RB41 for building facades, glassy DG41 for balustrades, and tri-layered QS41 for acoustic insulation. Other variations include RB11, an identical blend of RB41 except at half thickness (0.76 mm versus 0.38 mm). In all cases, the PVB polymers come with a given roughness that renders them opaque. The presence of vinyl alcohol group in the polymer, which can interact with water via hydrogen bonding, renders them highly hygroscopic meaning they readily absorb water from the surrounding atmosphere and must therefore be kept under controlled humidity conditions~\citep{arauz2023water, chen2023effect, botz2019experimental, desloir2019plasticization}.

We have measured the rheological properties of RB41, DG41, and QS41 using shear rheometry and found the blends to be viscoelastic in nature, i.e., they exhibit elastic and viscous responses that are well captured by a Maxwell (shear) relaxation modulus that is time-temperature dependent. In figures S1\textbf{A}-\textbf{C} in the supplementary material, we present the time-temperature state diagram for these polymers. In general, they may behave as glassy (low temperature/short timescales), rubbery (mid temperature/medium timescales), and as viscous melts (high temperatures/long timescales). To put the time/temperature scales into perspective, at 25°C RB41 is rubbery-like and would require $10^9$s ($\sim32$years) to reach the melt state. In contrast, at the temperature of lamination (e.g., 140°C), this time is only in the order of minutes!

\begin{figure*}[]
\includegraphics[width=0.7\textwidth]{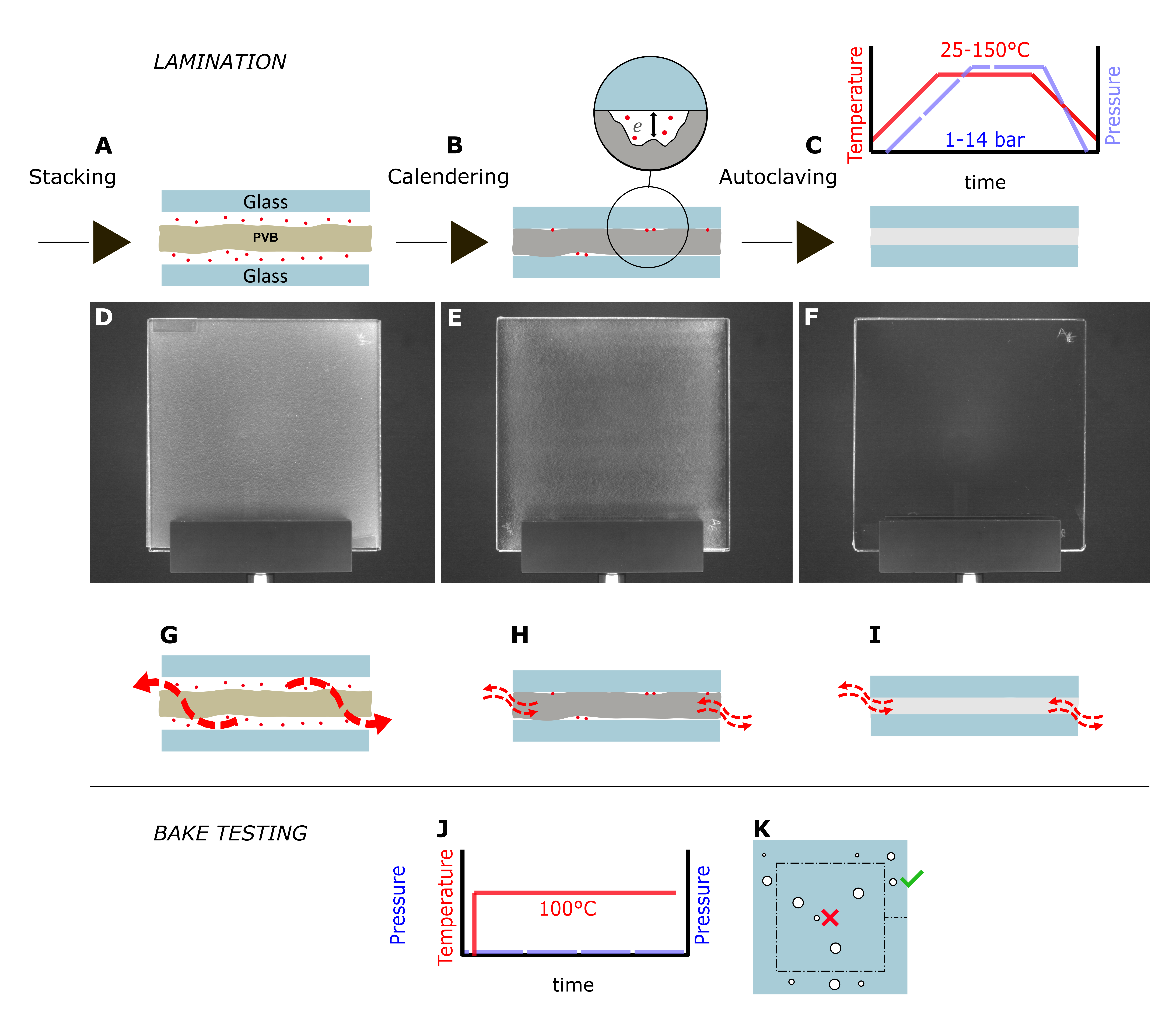}
\centering
\caption{\textbf{A-C} Glass lamination: stacking, calendering, and autoclaving. Red dots represent air molecules that become trapped in the polymer roughness $e$ at the polymer/glass interfaces. Across the steps, the nuanced shades of grey of the PVB polymer are meant to convey transparency transformations. \textbf{D-F} Images of laboratory-scale samples against a dark background that highlight how transparency evolves during the lamination process. The size of the samples is 10x10 cm. \textbf{G-I} Mass flux in the glass assembly. The size of the red arrows illustrates the relative speed with which gases can flow in or out of the sample. Fluxes in the z-direction are forbidden by the glass slides. \textbf{G} In the stacking step, gases can \textit{rapidly} flow, inwardly or outwardly, because the edges of the assembly are open and the polymer roughness forms a percolating network. The vertical distance between the glass and polymer layers has been exaggerated for artistic purposes. \textbf{H}, \textbf{I} After calendering and autoclaving, gases \textit{slowly} diffuse through the polymer bulk since the edges are sealed. \textbf{J}, \textbf{K} Bake test protocol and criteria for failure($\color{red}{\times}$)/success(\textcolor{green}{\checkmark}).} 
\label{fig2}
\end{figure*}

The lamination process involves three main steps  (\hyperref[fig2]{figures~2}A-C): stacking, calendering, and autoclaving. There is also the aforementioned quality control step, i.e., the bake test. Each step is scientifically rich enough to warrant its own investigation. To be expeditious, we do not discuss the peculiarities of the industrial process itself (to this end, see \citet{belis2019architectural} for an introductory discussion) but rather our own faithful laboratory scale replication. The latter allow us to draw scientific insights not accessible at the industrial scale. In particular, we circle the discussion around the many transformations the polymer/glass assembly undergoes in terms of optical appearance, mass transfer, adhesion, and polymer rheological states\textemdash all of which are related to bubble formation. 

\textbf{Stacking.} A PVB sheet of RB41 (roughness of $\sim40\mu$m) is stored under humidity-regulated air inside a climatic chamber to fix the amount of water dissolved in the bulk in a process known as \textit{conditioning}. The sorption process is self-similar with temperature and thereby simplifies the number of parameters which must be controlled to a single one: the chemical activity $a$ of water vapour in air~\citep{arauz2023water}. The standard conditions are $a=0.25$ which results in RB41 having $\sim 0.4\%$wt of water dissolved in the bulk. Then, a \textit{PVB sandwich} of glass/PVB/glass is stacked in a clean room.  Atmospheric air becomes trapped in the polymer roughness at both polymer/glass interfaces and the sandwich is thus opaque or hazy (\hyperref[fig2]{figure~2}D). There is no appreciable adhesion between the glass and the polymer (in the figure, the sample is held in place using tape) and gases can freely flow either in or out of the sandwich via the exposed polymer edges (\hyperref[fig2]{figure~2}G). In this regard, the polymer roughness forms a percolating network with the outside environment. The polymer can be considered as very much rubbery-like throughout this step. Finally, the stacking step hints at \textit{two gases} as the possible culprits behind bubble formation: \textcolor{blue}{\textit{water}} dissolved in the PVB bulk and \textcolor{red}{\textit{air}} trapped by the polymer surface roughness. 

\textbf{Calendering.} The PVB sandwich is slightly heated in a convection oven (e.g., 90°C for 15 mins) and bonded together using a nip-roll press (Linea DH-360, 1.4 m min\textsuperscript{-1}) forming what in the art is called a \textit{pre-press}. The PVB sandwich turns translucent as air is partially removed. At the same time, adhesion greatly develops, and the glass/polymer cannot be set apart. The edges of the sample become somewhat transparent because of perfect adhesion between the polymer and the glass (\hyperref[fig2]{figure~2}E).  The assembly is thus \textit{sealed} and gases may either only enter or leave \textit{through} the polymer bulk in a process that is bound by mass diffusion (\hyperref[fig2]{figure~2}H). The percolating network formed by the random polymer roughness is no longer present and gases are trapped in \textit{interfacial bubbles} or gas pockets of irregular shape (hereinafter, we shall refer to such bubbles as \textit{non-spherical} bubbles for clarity). Despite the increase in temperature, the polymer remains rubbery-like during the calendering step. From this step onwards, the non-spherical bubbles (or any other kind of bubble) form a closed thermodynamic system with the bulk of the PVB polymer, insofar as they are sufficiently away from the edges of the glass sample.  

\textbf{Autoclaving.} The pre-press is heated (e.g. 140°C) while simultaneously subjected to a hydrostatic load (e.g., 10-14 bar) (\hyperref[fig2]{figure~2}C).  Gaseous exchanges take place between the non-spherical bubbles and the polymer bulk (see $\S$~\ref{sec:phys_desc}). A typical autoclave cycle involves a complex schedule of heating and cooling ramps connected by an isothermal segment, hereinafter the \textit{hold} temperature. The pressure schedule is similarly intricate and may either lag, be in phase, or ahead of the temperature schedule. Post autoclaving, the pre-press is rendered transparent (\hyperref[fig2]{figure~2}F) thus yielding a finished sample of laminated safety glass, adhesion is maximized, while the edges remain sealed (\hyperref[fig2]{figure~2}I). The polymer, however, experiences a rubbery-to-viscous transition and back. Post autoclaving, the amount of water dissolved in the PVB bulk does not appreciably vary compared to the initial stacking step (see figure S3 in the supp. material). However, extra air is dissolved in the bulk from the non-spherical bubbles (see $\S$~\ref{sec:bake_test}). 

\textbf{Bake test.} A sample of LSG is heated for 16hrs at 100°C inside a pre-heated convection oven (Memmert) and is afterwards visually inspected for bubble formation (\hyperref[fig2]{figure~2}J). Two types of bubbles are considered (\hyperref[fig2]{figure~2}K):

\begin{itemize}
\item \textit{Edge bubbles} are found within 15mm from the edges. These bubbles are associated with mass diffusion and are thus neglected in the test.  We can estimate the lengthscale in question as $L\sim\sqrt{Dt}\sim 2-8$mm, where $D\sim1-10\times10^{-10}$ m\textsuperscript{2}/s is the representative diffusion coefficient and t is the test time.
\item \textit{Full face bubbles}  are  located deeper than 15mm inside the glass samples and constitute failure regardless of size and number.
\end{itemize}

\section{Toy Experiments}\label{sec:toy_exp}

\begin{figure*}[]
\includegraphics[scale=0.70]{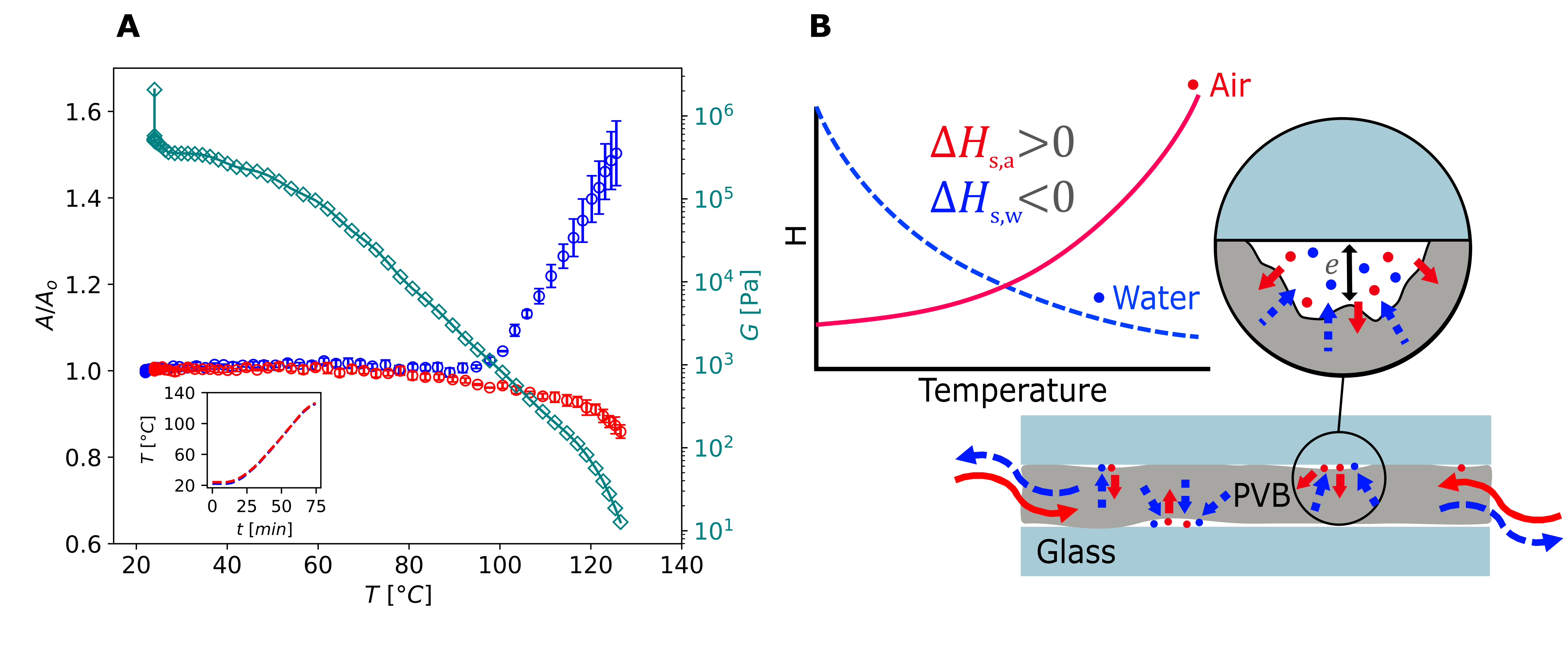}
\centering
\caption{\textbf{A} Example of bubble dynamics in a pre-press (RB41) during a toy autoclave schedule which included a heating ramp without overpressure. \textcolor{black}{\scalebox{1.2}{$\circ$}}  Interfacial bubble ratio $A/A_o$ (bubble area/initial bubble area) inside the target area in the sample. Two PVB humidity conditioning levels were tested, standard ($\color{blue}a=0.25$) and reduced humidity ($\color{red}a=0.05$). \textcolor{teal}{\scalebox{1.2}{$\diamond$}} PVB relaxation modulus using eq.~\ref{eq:Gen_Maxwell_TTS} for RB41 at standard humidity conditions only. The inset is the temperature schedule (color-matched to the main figure), \textcolor{blue}{\rule[.5ex]{1em}{.5pt}} \textcolor{blue}{\rule[.5ex]{1em}{.5pt}}, \textcolor{red}{\rule[.5ex]{1em}{.5pt}} \textcolor{red}{\rule[.5ex]{1em}{.5pt}}. See figure S4 in the supplementary material for the experimental set-up details. \textbf{B} Sketch of gaseous exchanges that take place in a pre-press when heated: \textcolor{blue}{\textit{water}} escapes from the bulk towards the non-spherical bubbles seeking to inflate them, while \textcolor{red}{\textit{air}} does the opposite. Each gas has a distinct heat of solution $\Delta H_s$ sign which leads them to follow distinct thermodynamic paths with temperature, solubility ($H$) wise, in the PVB polymer. For completeness, the sketch also depicts the gaseous exchanges that take place between the PVB polymer and the surrounding amtosphere via the exposed edges of the polymer. Air \textit{enters} while water \textit{leaves}.} 
\label{fig3}
\end{figure*}

We developed a set of transparent autoclaves that allowed us to visualize and track bubbles in whatever type of glass sample we desired (see figure S4 in the supplementary material). The autoclaves are flexible enough to accommodate arbitrary schedules of temperature and pressure. In \hyperref[fig3]{figure~3}A, we present a simple, yet powerful toy experiment that elegantly showcases the various physical ingredients behind bubble growth in glass assemblies with PVB. In said figure, a pre-press was subjected to an isobaric heating ramp and the non-spherical bubbles were tracked via the \textit{interfacial bubble ratio} $A/A_o$ (normalized surface area of non-spherical bubbles in the target area of the sample). Two PVB conditioning levels were probed: standard \textcolor{blue}{\textit{moist}} ($a=0.25$) and \textcolor{red}{\textit{dry}} PVB (reduced moisture content, $a=0.05$).

In the pre-press with moist PVB (blue curve), the non-spherical bubbles initially remain stable during the early stages of heating ($A/A_o\sim 1$) up to 80-90°C whence explosive growth ensued ($A/A_o>1$). We thus rapidly conclude that the natural tendency of a glass assembly at elevated temperatures with standard PVB is to bubble! New bubbles never nucleate, and coalescence is likewise not observed.

The growth of the non-spherical bubbles during the heating ramp betrays the volatile essence of water in PVB. This gas has a negative heat of solution in the polymer (\textcolor{blue}{$\Delta H_{s,w}$}$<0$) and therefore, has a solubility which diminishes when the temperature is increased~\citep{arauz2023water}. Accordingly, water tends to escape from the PVB bulk and form bubbles (the reverse evidently holds under a cooling ramp). The more water is initially dissolved in the bulk, the larger the bubbles (not shown). 

Repeating the toy experiment with the pre-press using dry PVB (red curve), on the contrary, leads to no bubbling at all. In this case, bubbles shrink rather than grow during the heating ramp as air\textemdash the predominant gas in the non-spherical bubble (see $\S$~\ref{sec:phys_desc})\textemdash becomes more soluble in PVB with temperature ($\color{red}{\Delta H_{s,a}}>0$)~\citep{arauz2025champagne}. In other words, air flows from the non-spherical bubbles towards the polymer bulk when the later is heated.

Therefore, thermodynamically speaking and as sketched in \hyperref[fig3]{figure~3}B, water promotes non-spherical bubble growth, while air favors bubble shrinkage in pre-press assemblies during a heating ramp. In other words, the gases compete for the fate of these bubbles. Notice, however, that below a seemingly critical temperature, the competition is rendered moot as the bubbles are stable ($A/A_o=cte$). As we will show in the next section, this stability is linked to the viscoelastic nature of the polymer\textemdash in particular its dramatic relaxation with temperature (see the green curve in \hyperref[fig3]{figure~3}B). Only when the PVB polymer softens sufficiently enough can bubbles grow or shrink.

So far, our results indicate a strong link between water and bubble formation in glass assemblies with PVB, specially in a pre-press during autoclaving at elevated temperatures. Is thus water directly responsible for bubbles in finished samples of LSG? if so, this begs the question, what is the role played by air?  We answer these questions in the remaining sections of the paper. 

\section{Physical Description}\label{sec:phys_desc}

We propose hereinbelow a physical model for the growth of bubbles in finished assemblies of laminated safety glass with PVB but which is general enough to cover other time-temperature dependent soft matter materials. The model determines the bubble radius $R$ from a system of coupled linear differential equations that simultaneously address mass transport of multiple gases, thermal effects, and rheological phenomena. We make some simplifying assumptions:  

\begin{enumerate}[i]
\item There is a pre-existent gas nucleus of radius $R_o$ of a micrometric size (see $\S$~\ref{sec:bake_test}). 

\item Confinement effects are negligible, at least at first order. For simplicity, we assume spherical symmetry throughout.

\item The medium is taken as infinite, thus, there is ample gas supply.

\item Bubbles are well-separated from one another so that bubble-to-bubble effects are ignored.

\item The temperature field is homogeneous in time and space. 

\item For mass transport specifically,

\begin{enumerate}
\item There are only two gaseous species of relevance, \textcolor{blue}{\textit{water}} and \textcolor{red}{\textit{air}}, each one behaving as ideal gases. Air, being a mixture of several gases, is approximated as Nitrogen for simplicity\footnote{Of the multiple gases that constitute atmospheric air, only nitrogen's mass transport properties (e.g., solubility constant, diffusion coefficient) have been characterized for different temperatures and pressures in the PVB polymer\cite{arauz2025champagne}. Therefore, assumption vi.a is in principle utilitarian but ultimately tenable since air is largely a two-component mixture of N\textsubscript{2} and O\textsubscript{2}. While the PVB polymer has been shown to absorb less nitrogen than oxygen at room conditions under equal pressure (ratio of nitrogen to oxygen equal to $r^{PVB}_{N_2/O_2}=0.67$)\citep{haraya1992permeation}, nitrogen is by far the dominant gas, both in the bubbles and in dissolved form in the PVB bulk, given its prevalence in atmospheric air ($r^{air}_{N_2/O_2}=3.76$).}.
\item Each species has a dissolved concentration profile in the PVB bulk that obeys the diffusion equation separately.
\item The time dynamics are quasi-static.
\item Transport constants, in particular the activation energy for diffusion $E_d$ and the heat of solution $\Delta H_s$ are constant with temperature and pressure. 
\end{enumerate}
\item The deformation (and its rate) is small, i.e., the PVB polymer stays in the range of linear viscoelasticity.
\end{enumerate}

\textbf{Mass Transport}. In broad strokes, we solved the diffusion equation to obtain the concentration of dissolved gas in the polymer and thence determined the flux of gas at the bubble interface from Fick’s law. Expressions for the bubble composition were then obtained from the mol fraction and Dalton’s law of partial pressures. 

The diffusion equation reads

\begin{equation} \label{eq:diff_equation}
\frac{\partial c_j}{\partial t}=D_j\Delta c_j
\end{equation}

where $c_j, D_j$ are the gas concentration profile and diffusion coefficient of gas $j$. Under i-iv,vi.b the boundary conditions for each gas are identical to those proposed by \citet{epstein1950stability} for a mono-component bubble of radius $R$ embedded in an infinite medium

\begin{equation}
\begin{split}
c_j(r,0)&=c_{i,j}, \ r>R \\ 
c_j(r,t)&=c_{i,j}, \ r\to\infty, \ t>0 \\ 
c_j(R, t)&=c_{s,j}, \ r=R, \ t>0,
\end{split}
\label{eq:bound_cond}
\end{equation}

where $c_{i,j}$ is taken as the homogeneous initial concentration of dissolved gas in the hosting medium and $c_{s,j}=H_jp_j$ is the gas concentration at the bubble interface which we assume is given by Henry's law for a gas with solubility $H_j$ and partial pressure $p_j$. We suppose that $c_{s,j}$ is achieved \textit{instantaneously} because the gas concentration in the bubble, as set by the gas density, is undoubtedly much larger than the concentration of dissolved gas in the polymer bulk. Therefore, the bubble interface reaches equilibrium much faster compared to the timescale set by diffusion. Thence, solving eq.~\ref{eq:diff_equation} under the boundary conditions in eq.~\ref{eq:bound_cond} leads to

\begin{equation} \label{eq:c_gradient}
\bigg(\frac{\partial c_j}{\partial r}\bigg)_{r=R}=-(c_{s,j}-c_{i,j}) \bigg[\frac{1}{R}+\frac{1}{\sqrt{\pi D_j t}}
\bigg].
\end{equation}

The above is simplified by noting that the rightmost term vanishes with time whenever $1\gg R/\sqrt{\pi D t}$, which is our case. We thus have

\begin{equation} \label{eq:c_gradient_steady}
\bigg(\frac{\partial c_j}{\partial r}\bigg)_{r=R}=\frac{-(c_{s,j}-c_{i,j}) }{R}, 
\end{equation}

and the flux is 

\begin{equation} \label{eq:Fick's_1st_law}
J_{j, r=R}=-D_j\nabla c_{j, r=R}.
\end{equation}

Merging eqs.~\ref{eq:c_gradient_steady},~\ref{eq:Fick's_1st_law}, defining flow \textit{into} the bubble as positive, and multiplying the result by the bubble's surface area yields

\begin{equation} \label{eq:gasj-moles}
\frac{dn_j}{dt}=-4\pi R D_j (c_{s,j}-c_{i,j}).
\end{equation}

where $dn_j/dt$ is the variation of moles for gas $j$ with time.

Eq.~\ref{eq:gasj-moles} predicts mass exchanges (or lack thereof) between the hosting medium and the bubble because of oversaturation ($c_{s,j}-c_{i,j}<0$), saturation ($c_{s,j}-c_{i,j}=0$), and undersaturation ($c_{s,j}-c_{i,j}>0$). These conditions are likewise quantified via the oversaturation driver $f_j=c_{i,j}/c_{s,j}$. Accordingly, we have $f_j>1$ (gas flow into the bubble), $f_j=1$ (no mass flow) and $f_j<1$ (flow away from the bubble).

We now write explicit expressions for air and water (vi.a)

\begin{equation} \label{eq:gas1-moles}
{\color{red}\frac{dn_a}{dt}}=-4\pi R {\color{red}D_a }({\color{red}c_{s,a}-c_{i,a}})
\end{equation}

\begin{equation} \label{eq:gas2-moles}
{\color{blue}\frac{dn_w}{dt}}=-4\pi R {\color{blue}D_w}({\color{blue}c_{s,w}-c_{i,w}}),
\end{equation}

and the total number of moles in the bubble $n_B$ is simply the aggregate of the two

\begin{equation} \label{eq:bubble-moles}
\frac{dn_B}{dt}={\color{red}\frac{dn_a}{dt}}+{\color{blue}\frac{dn_w}{dt}}.
\end{equation}

The bubble composition is obtained from the mol fraction of the gases $x_j=n_j/n_B$, such that, $\sum x_j=1$. Therefore, 
\begin{equation} \label{eq:fraction_air}
{\color{red}\frac{dx_a}{dt}}=\frac{1}{n_B}\bigg({\color{red}\frac{dn_a}{dt}}-\frac{\color{red}n_a}{n_B}\frac{dn_B}{dt}\bigg)
\end{equation}

\begin{equation} \label{eq:fraction_water}
{\color{blue}\frac{dx_w}{dt}}=-{\color{red}\frac{dx_a}{dt}}.
\end{equation} 

Since we take the gases as ideal (vi.a), Dalton’s law links simultaneously the partial pressure of the gases, the mol fractions and the bubble pressure $p_j=x_jP_B$. Accordingly, for each gas we have

\begin{equation} \label{eq:pressure_air}
{\color{red}\frac{dp_a}{dt}}={\color{red}x_a} \frac{dP_B}{dt}+P_B {\color{red}\frac{dx_a}{dt}}
\end{equation}

\begin{equation} \label{eq:pressure_water}
{\color{blue}\frac{dp_w}{dt}}={\color{blue}x_w} \frac{dP_B}{dt}+P_B {\color{blue}\frac{dx_w}{dt}}.
\end{equation}

Eqs.~\ref{eq:gas1-moles} 
-~\ref{eq:pressure_water} are general in nature and can be used to solve the mass transport problem for any two-component bubble system. They can also be expanded to n-component bubbles by slightly modifying eq.~\ref{eq:fraction_water} to resemble eq.~\ref{eq:fraction_air} for n gases. 

\textbf{Thermal effects}. Changes in temperature modify transport constants as well as material properties in important and non-trivial ways.  

We suppose diffusivity and solubility are temperature activated processes that follow an Arrhenius behaviour with temperature. In compact form, both transport mechanisms can be expressed as

\begin{equation}
X_j=X_{j,r}\exp\bigg({\frac{-Y_{j}}{R_u}\bigg[\frac{1}{T}-\frac{1}{T_{r}}\bigg]}\bigg)
\label{eq:transport_constants}
\end{equation}

where $R_u$ is the universal gas constant, $X_j$ is either the solubility $H_j$ or diffusivity $D_j$, $X_{j,r}$ is the respective constant at an arbitrary reference temperature $T_r$, and $Y_j$ is the corresponding heat of solution $\Delta H_{s,j}$ or the activation energy of diffusion $E_{d,j}$. The former dictates whether gas solubility increases ($\Delta H_s>0$) or decreases ($\Delta H_s<0$) with temperature. The latter is always positive and controls how fast molecules move through the polymer matrix.  

The initial gas concentration in the PVB bulk (during conditioning) is set by the partial pressures of water and air in the surrounding atmosphere. As a result,  
\begin{align}
\textcolor{blue}{c_{i,w}}&=\textcolor{blue}{H_{w}(T_i)}\textcolor{blue}{p_{i,w}}\\
\textcolor{red}{c_{i,a}}&=\textcolor{red}{H_{a}(T_i)}\textcolor{red}{p_{i,a}}
\label{eq:air_conc}
\end{align}

and,

\begin{align}
\textcolor{blue}{p_{i,w}}&=a P_{Sat}(T_i)\label{eq:initial_air_p}\\
\textcolor{red}{p_{i,a}}&=P_o-a P_{Sat}(T_i)\label{eq:initial_water_p}.
\end{align}

where $P_{Sat}$ is the vapor pressure of water, $T_i$ is the initial experimental temperature, and $P_o$ is the reference atmospheric pressure (101.325 kPa)

The above conditions set the initial composition of the bubbles in the pre-press since chemical equilibrium mandates $f_j=1$ for the gases inside these bubbles and 
those in the polymer bulk. We accordingly have $\textcolor{red}{x_{i,a}}\sim \textcolor{red}{p_{i,a}}/P_o\sim0.99$ and $\textcolor{blue}{x_{i,w}}\sim \textcolor{blue}{p_{i,w}}/P_o\sim0.01$. During autoclaving, the equilibrium is shifted so that $f_j\neq1$ and hence, mass exchanges take place between the non-spherical bubbles in the pre-press and the polymer bulk. For a finished assembly of glass, which is our goal, the initial conditions are slightly shifted (see $\S$~\ref{sec:bake_test}). Moreover, for most structural glass applications, \textit{chemical equilibrium is broken solely by temperature} since pressure can be assumed as constant.  

\begin{figure*}[htp]
\includegraphics[width=0.7\textwidth]{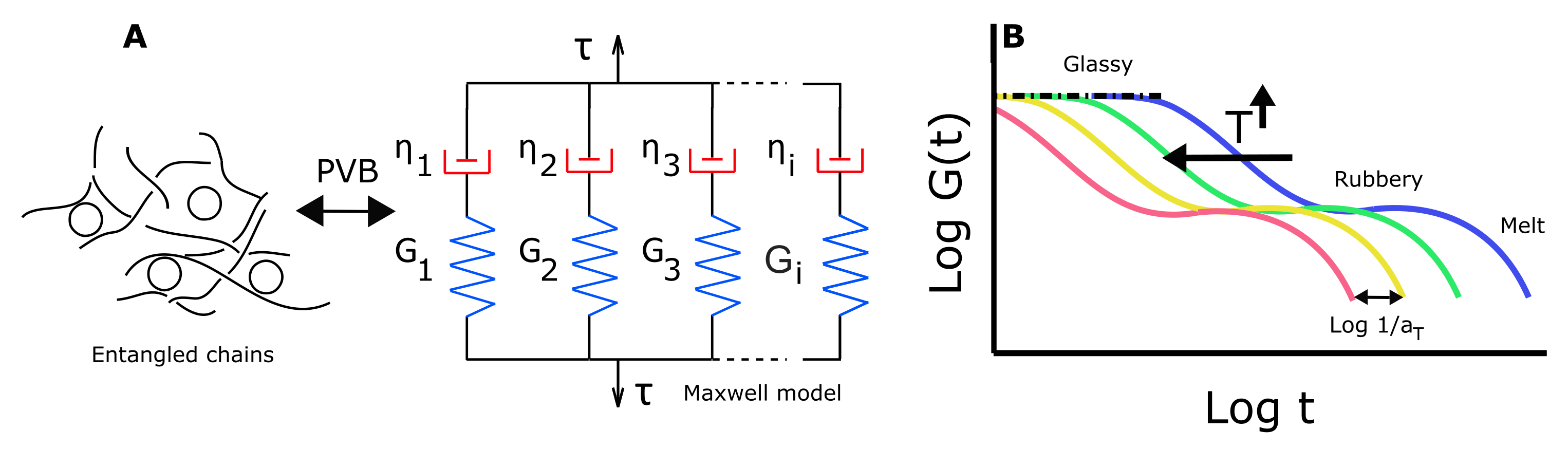}
\centering
\caption{\textbf{A} The PVB polymer is thought of as entangled chains separated by a plasticizer molecule (circles). The corresponding physical description is a Generalized Maxwell model which includes a plurality of parallel branches of Newtonian dashpots of viscosity $\eta_i$ and Hookean springs with elasticity $G_i$ to model the simultaneous viscous and elastic behavior of the polymer. Both features are compactly captured via a shear relaxation \textcolor{teal}{$G(t,T)$} that is time-temperature dependent. \textbf{B} Sketch of PVB relaxation with time and temperature in the log(t) space. The curves are self-similar with temperature and differ only via a shift factor $a_T$. The overall relaxation \textcolor{teal}{G(t)} includes three distinct polymer states: glassy (cold temperatures/short timescales), rubbery (mid temperatures/mid timescales), and viscous melt (high temperatures/long timescales).} 
\label{fig4}
\end{figure*}

\textbf{Rheology.} The presence of rheological effects adds an extra stress at the bubble interface and thus affects the overall bubble pressure. Generally speaking, under spherical symmetry and in the limit of small deformation (vii), $P_B$ is~\citep{fyrillas2000factors,kloek2001effect}

\begin{equation} \label{eq:press_bubble}
P_B=P_o-\tau_{rr}\vert_{r=R}
\end{equation}

where $\tau_{rr}\vert_{r=R}$ is given by the radial component of the stress tensor evaluated at the vicinity of the bubble. 

For PVB, the extra stress is related to the macroscopic strain $\gamma_r$ via the shear relaxation modulus \textcolor{teal}{$G$},

\begin{equation}
\tau_{rr}=2{\color{teal}G(t,T)}\gamma_r
\label{eq:tau_rr_model}
\end{equation}

with \textcolor{teal}{$G$} in the linear regime being given by a generalized Maxwell model comprised by a plurality of parallel elements of Hookean springs and Newtonian dashpots connected in series (\hyperref[fig4]{figure~4}A). The springs account for the elastic response of the polymer, while the dashpots address viscous dissipation. Each spring/dashpot element has a relaxation time $\lambda_i=\eta_i/G_i$. Since relaxation curves for PVB at different temperatures are self-similar and differ only by a shift factor $log (1/a_T)$ in the time space (\hyperref[fig4]{figure~4}B), all relaxation times decay identically with temperature. We can thus build a time-temperature dependent Maxwell model~\citep{arauz2022extended} 

\begin{equation}
\textcolor{teal}{G(t,T)}=\sum_i G_{i} \times exp \bigg(\frac{-t}{a_T(T) \times \lambda_i}\bigg)
\label{eq:Gen_Maxwell_TTS}
\end{equation}

where $i=1...10$, and the shift factors $a_T$ at any temperature $T$ can be modeled analytically using the well-known WLF law of thermorheological simple materials

\begin{equation}
log(a_T)=\frac{-C_1(T-T^*)}{C_2+T-T^*} 
\label{eq:WLF_law}
\end{equation}

and $C_1, C_2$ are experimental constants at an arbitrary reference temperature $T^*$~\citep{arauz2022extended} (see table 1 in the supp. material for the complete set constants and figure S2 for a comparison of the model to experimental data).  

Finally, the macroscopic strain $\gamma_r$, at an arbitrary radial location $r$, is obtained from mass continuity by integrating the rate of deformation imposed by a growing bubble in an infinite medium~\citep{yang2005model,webb2011effect}

\begin{equation}
\gamma_r=\frac{-2}{3}\bigg(\frac{R^3-R_o^3}{r^3}\bigg). 
\label{eq:mac_strain}
\end{equation}

Combining eqs.~\ref{eq:press_bubble},~\ref{eq:tau_rr_model}, and ~\ref{eq:mac_strain} yields,

\begin{equation}
P_B=P_o+\frac{4}{3}{\color{teal}G(t,T)}\bigg(1-\frac{R_o^3}{R^3}\bigg),
\label{eq:bubble_maxwell}
\end{equation}

whence we find,

\begin{equation} \label{eq:diff_bubble_press}
\frac{dP_B}{dt}=\frac{4\color{teal}G(t,T)}{R}\bigg(\frac{R_o}{R}\bigg)^3\frac{dR}{dt},
\end{equation}
where the term $dG/dt$ was negligible in our case. Indeed, including it in our numerical model led to no appreciable differences in our results. 

\textbf{Bubble Radius}. The radius differential $dR/dt$ is obtained from the ideal gas law, which for general purposes we write in differential form

\begin{equation}\label{eq:diff_ideal_gas}
\frac{dR}{dt}=\frac{1}{4\pi R^2}\frac{R_u}{P_B}\bigg(n_B\frac{dT}{dt}+T\frac{dn_B}{dt}-\frac{n_BT}{P_B}\frac{dP_B}{dt}\bigg)
\end{equation}

The temperature field is typically imposed, ergo the temperature (and its possible variation) in the vicinity of the bubble is a known quantity (v). Combining eqs.~\ref{eq:diff_bubble_press},~\ref{eq:diff_ideal_gas} leads to

\begin{equation} \label{eq:bubble_radius_model}
\frac{dR}{dt}=\frac{1}{4\pi R^2 }\frac{R_u}{P_o+4/3{\color{teal}G(t,T)}} \bigg(n_B\frac{dT}{dt}+T\frac{dn_B}{dt}\bigg).
\end{equation}

Eq.~\ref{eq:bubble_radius_model} is very general in its nature. It can be used to model bubble growth (or shrinkage) under isothermal or non-isothermal conditions, can account for multiple gases via the mass transport problem, and can even be used to simulate classical Neo-Hookean solids (under small deformation) if \textcolor{teal}{$G(t,T)$} is set as a constant. For a medium with a different rheology, it suffices to determine $\tau_{rr}$ via a constitutive equation and replace eq.~\ref{eq:bubble_radius_model} by eq.~\ref{eq:diff_ideal_gas}.  

\begin{figure*}[h!t!]
\includegraphics[width=0.7\textwidth]{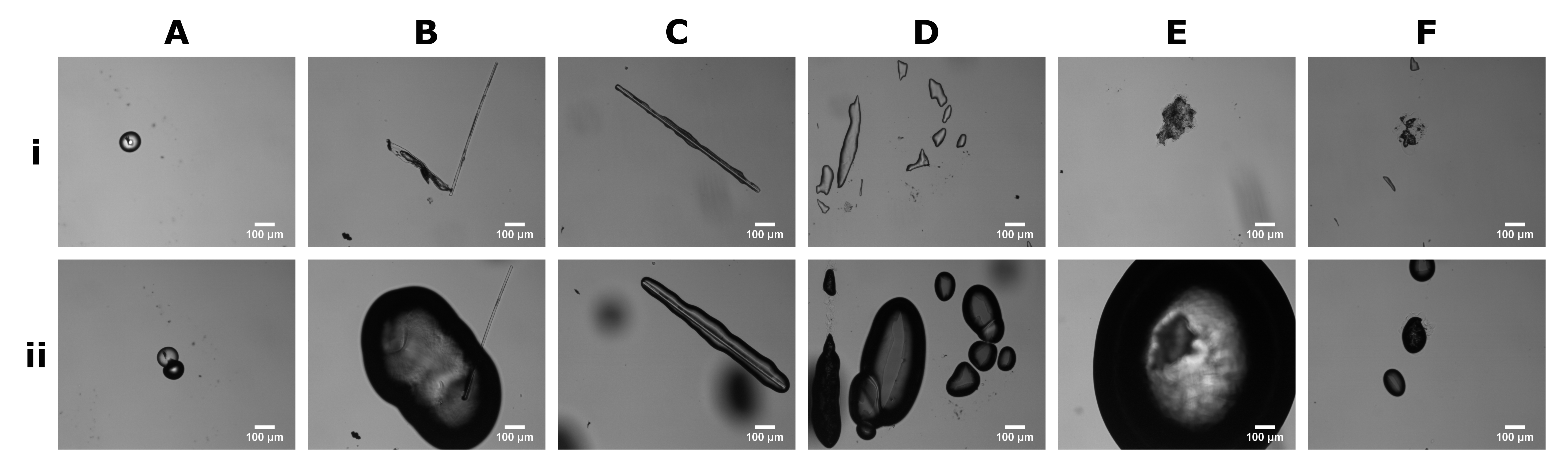}
\centering
\caption{Gas nuclei before (row i) and after (row ii) bake testing at 100°C for 16hrs.} 
\label{fig5}
\end{figure*}

Likewise, Eq.~\ref{eq:bubble_radius_model} already explains, at least qualitatively, some of the bubble observations made in $\S$~\ref{sec:toy_exp}. At low temperatures, the polymer elasticity is quite high ($G\sim 10^7$Pa). It must then be the case that under these circumstances, the $\color{teal}{G(t,T)}$ term in the numerator dominates the dynamics, which results in bubbles being stabilized $dR/dt\sim 0$ even under a non-isothermal temperature field. As time progresses and/or the temperature is increased, this term continuously loses in strength until it is overcome by the combined effect of thermal dilation $dT/dt$ (which inherently tends to swell the bubble) and mass transport $dn_B/dt$. An upper bound for this condition is $P_o\gg 4/3\color{teal}{G}$, i.e., when the polymer rheology effectively vanishes and no longer affects the bubble pressure. The actual threshold may be lower than this depending on the magnitude of the driver and the temperature. 

All things equal, the mass transfer term favors bubble growth, stability, or shrinkage depending on whether the gas is under oversaturation, saturation, or undersaturation conditions. When dealing with multiple gases, the situation is intricate as gases may cooperate or compete against one another for the fate of the bubble. This is indeed the case for non-spherical bubbles in the pre-press. Air favors bubble shrinkage while water promotes bubble growth (\hyperref[fig3]{figure~3}B) because these gases have different heat of solution signs in PVB. In a finished sample of LSG, the situation is different, and the gases cooperate to induce bubble growth (see $\S$~\ref{sec:bake_test}). Finally, we remark that under isothermal conditions ($dT/dt=0$) or after a temperature jump, the absolute value of $T$ plays a role by directly influencing how fast and how large bubbles may become. As we discuss later in the paper, the situation is a bit more involved as there is also a \textit{size effect} at play.

\section{Nucleation}\label{sec:nucleation}

Using our transparent autoclave, we investigated different lamination parameters deemed a priori conducive to bake test failure, such as PVB type and moisture conditioning level as well as autoclave temperature and pressure conditions. The common denominator for bake test failure was often the presence of undissolved gaseous inclusions that served as nucleation sites in the glass assembly. These sites are invisible to the unaided human eye but were relatively large enough to cause a distinction in the images taken by our autoclave set-up (resolution $\sim10\mu$m). For samples that failed the bake test, the autoclave images often appeared \textit{grainy} or \textit{dusty} because they were populated by pixel-sized dark dots (figure S5\textbf{A} in the supplementary material). The latter can easily pass off as dust-like contamination.  After bake testing and in numerous occasions, bubbles grew in many of these pixel-sized dots (see figure S5\textbf{B} in the suppl. mat.).

We thus proceed to carefully scan our samples before and after bake testing under the microscope. In many of the aforementioned pixel-sized dots, we often found gas nuclei of different sizes and morphologies. They were always at one of the glass/polymer interfaces but never in the PVB bulk, i.e., nucleation was entirely of the heterogenous kind.  In \hyperref[fig5]{figure~5}, we present a small subset of the type of \textit{naturally} occurring nuclei we observed (row i, Cols. A – F) accompanied by the resulting bubbles that grew after bake testing (row ii, idem. Cols.). \hyperref[fig5]{Figure~5}A shows a rare nuclei variety, a seemingly spherical nuclei that led to a binary bubble system post bake testing. \hyperref[fig5]{Figure~5}B illustrates a gas inclusion \textit{inside} in what is apparently a fiber in a situation reminiscent of Champagne (see \citet{liger2008recent} for a review on the subject), while \hyperref[fig5]{figure~5}C is a glass inclusion \textit{around} a fiber. In \hyperref[fig5]{figure~5}D, undissolved non-spherical bubbles grow into large bubbles post bake testing. The bubbles apparently grew with a pinned contact line. See how the original contour of the nuclei is very much noticeable in the final bubble. \hyperref[fig5]{Figure~5}E is apparently a region of wrinkled or folded PVB matter that trapped gases. Finally, \hyperref[fig5]{figure~5}F is an \textit{artificially} made inclusion. A crenel was etched on the glass surface (towards the center of the image) using a UV laser. During autoclaving, the crenel was partially filled by PVB leaving behind undissolved gases whence a bubble grew during bake testing. The final size of the bubbles was proportional to the original volume of gas in the crenel (not shown). 

\begin{figure*}[h!t!]
\includegraphics[width=0.7\textwidth]{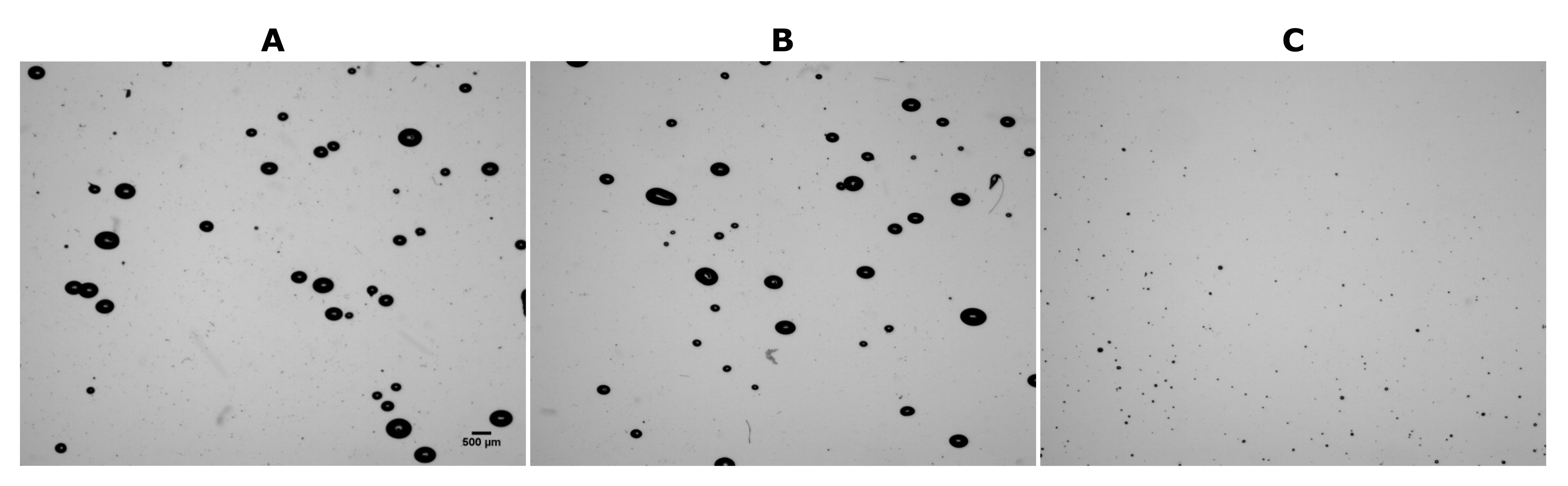}
\centering
\caption{Bake testing results for different PVB interlayers. \textbf{A} RB41 (standard). \textbf{B} QS41 (acoustic). \textbf{C} DG41 (stiff).} 
\label{fig6}
\end{figure*}

We close this section by briefly discussing the role played by interlayer rheology. To this end, \hyperref[fig6]{figures~6}A-C compare the bubbling behavior of RB41, QS41, and DG41 post bake testing respectively. While the commonality for failure was again the presence of undissolved gas nuclei, DG41\textemdash the stiffer version of the three PVB interlayers\textemdash seemed more resilient to test failure as it yielded comparatively smaller bubbles. This is probably a reflection of the high stiffness of this interlayer. On the other hand, RB41 and QS41 behaved roughly the same as demonstrated by their comparatively similar number of bubbles and their size.

\section{Describing the Bake Test}\label{sec:bake_test}

We used the physical description presented in $\S$~\ref{sec:phys_desc} to simulate the growth of bubbles during bake testing of finished assemblies of LSG. Specifically, we simultaneously solved the equations for the bubble radius (eq.~\ref{eq:bubble_radius_model}), mass transport (eqs.~\ref{eq:gas1-moles}   
-\ref{eq:pressure_water}), and bubble pressure (eq.~\ref{eq:diff_bubble_press}). The temperature field was assumed homogeneous throughout and in the form of a step jump from $T_i=20$°C to $T_{BT}=100$°C. Solubility and diffusivity were computed via eq.~\ref{eq:transport_constants}, and the polymer relaxation via eqs.~\ref{eq:Gen_Maxwell_TTS},~\ref{eq:WLF_law}. In total, our description requires more than 30 different physical constants that were determined by direct experimentation. See the suppl. material for the constant values.

Computationally, a constant step size of $\Delta t=0.1s$ was used and the total simulation time was equal to 16hrs. The variables were initialized and the system of equations was solved in a loop using the \textit{\textbf{odeint}} package in python. To simulate the bake test, two crucial assumptions were made:

\textbf{Oversaturation}. Post lamination, the amount of water in the PVB bulk remains unchanged (see figure S3 in the supp. mat.), but extra air is dissolved therein. To determine air oversaturation, we assume next ideal calendering, i.e., air is displaced but not absorbed in the PVB bulk. The edges of the glass assembly are likewise perfectly sealed so the high-pressure air applied during autoclaving cannot oversaturate the PVB bulk (except at the edges which we disregard). Under this framework, the total amount of air \textcolor{red}{$c^*_a$}, from the pre-press to the finished assembly, remains constant and is equal to

\begin{equation}
\textcolor{red}{c_a^*}=\textcolor{red}{c_{i,a}}+A\frac{e}{h}\textcolor{red}{\rho_a}(T_i),  
\label{eq:vol_air}
\end{equation}

where the left term is the amount of air dissolved in the PVB bulk from saturation conditions (eq.~\ref{eq:air_conc}) and the right term is the quantity of air present in the interfacial non-spherical bubbles. $A$ is the (initial) fraction of interfacial bubbles in the pre-press ($\approx 0.32-0.40)$, $e$ is the PVB surface roughness ($\sim 40\mu$m), $h$ is the PVB half-thickness, and \textcolor{red}{$\rho_a$} is the dry air density.

Eq.~\ref{eq:vol_air} can be re-written as 
\begin{equation}
\textcolor{red}{c_a^*}=H_a^*\textcolor{red}{p_{i,a}}
\end{equation}

where $H_a^*=$\textcolor{red}{$H_a$}($T_i$)$+A\frac{e}{h}\frac{\textcolor{red}{\rho_a}(T_i)}{\textcolor{red}{p_{i,a}}}$ is an effective solubility constant that takes into account the total number of air moles in the system. 

\textbf{Nucleus composition}. There is a nucleus of size $R_o$ of micrometric size, in mechanical equilibrium with the surroundings at atmospheric conditions so that at $t=0$, $P_B=P_o$ in agreement with eq.~\ref{eq:bubble_maxwell}, and is also in chemical equilibrium with the gases dissolved in PVB bulk. Therefore, the gaseous composition of this nucleus is the same as the non-spherical bubbles in a pre-press (eqs.~\ref{eq:initial_air_p},~\ref{eq:initial_water_p}), i.e., the initial partial pressure of our bubbles/nuclei remains unchanged after lamination.

With these assumptions, the driver for air in our experiments is \textcolor{red}{$f_a$}$\sim1.68-1.85$ at room conditions. We therefore reach a peculiar conclusion for finished assemblies of LSG: the gas nuclei contained therein are in a metastable state for air wants to come out of solution from the PVB bulk and hence, grow a bubble. It is only the (high) stiffness of the interlayer which prevents air from doing so at room conditions as already discussed in connection with eq.~\ref{eq:bubble_radius_model}. Note, however, how temperature is a key factor for bubble growth because it simultaneously (i) accelerates the polymer relaxation, thus progressively easing the constraints which prevent the nuclei from growing in the first place and (ii) shifts the chemical equilibrium of water vapor, which has bubble-forming tendencies to begin with in PVB.   

\section{When Water is \textit{Partly} to Blame}\label{sec:water_partly}

We prepared samples of RB41 at standard humidity conditions ($a=0.25$) and with a low quantity of initial interfacial bubbles ($A/A_o\sim0.4$). Under these conditions, the instantaneous oversaturation driver for water at 100°C is roughly $\textcolor{blue}{f_w}\sim40$, meaning that, like its air counterpart, water-driven bubbles are thermodynamically favorable during the bake test. Nevertheless, visible bubbles should only form provided that the original gaseous inclusion or nuclei are sufficiently large, i.e., there is a \textit{size effect}, which we discuss next.

\citet{kloek2001effect} have shown that bulk elasticity can arrest bubble dissolution, with the final bubble size $R_f$ being a function of the initial bubble radius $R_0$. In particular, larger bubbles shrink less compared to smaller ones. Furthermore, provided the elastic shear modulus is sufficiently high, the bubble may remain stable at $R_0$ regardless of initial size, i.e., there is a complex interplay between rheology and $R_o$ which ultimately dictates the final radius of the bubble. Our case is somewhat the reverse: larger values of $R_0$ lead to an equally greater bubble size $R_f$ after bake testing. Our point of departure, however, is not whether bubbles reach a final equilibrium but whether they become observable at all post bake testing. Rheology aside, the final bubble size is also inherently affected by the bubble surface area $4\pi R^2$ which governs the mass transfer problem. Evidently, larger bubbles have an equally greater surface area through which gas can diffuse. 

We have estimated the critical size that leads to an observable bubble post bake testing by performing a numerical sweep for nuclei sizes ranging between $R_0=10-100\mu m$. For reference, the resolution of the human eye is around $100\mu$m (the thickness of a hair strand). We took an arbitrary bubble size of half a millimeter as the threshold for a visible bubble, and thus, bake test failure.

\begin{figure*}[htp]
\includegraphics[width=0.54\textwidth]{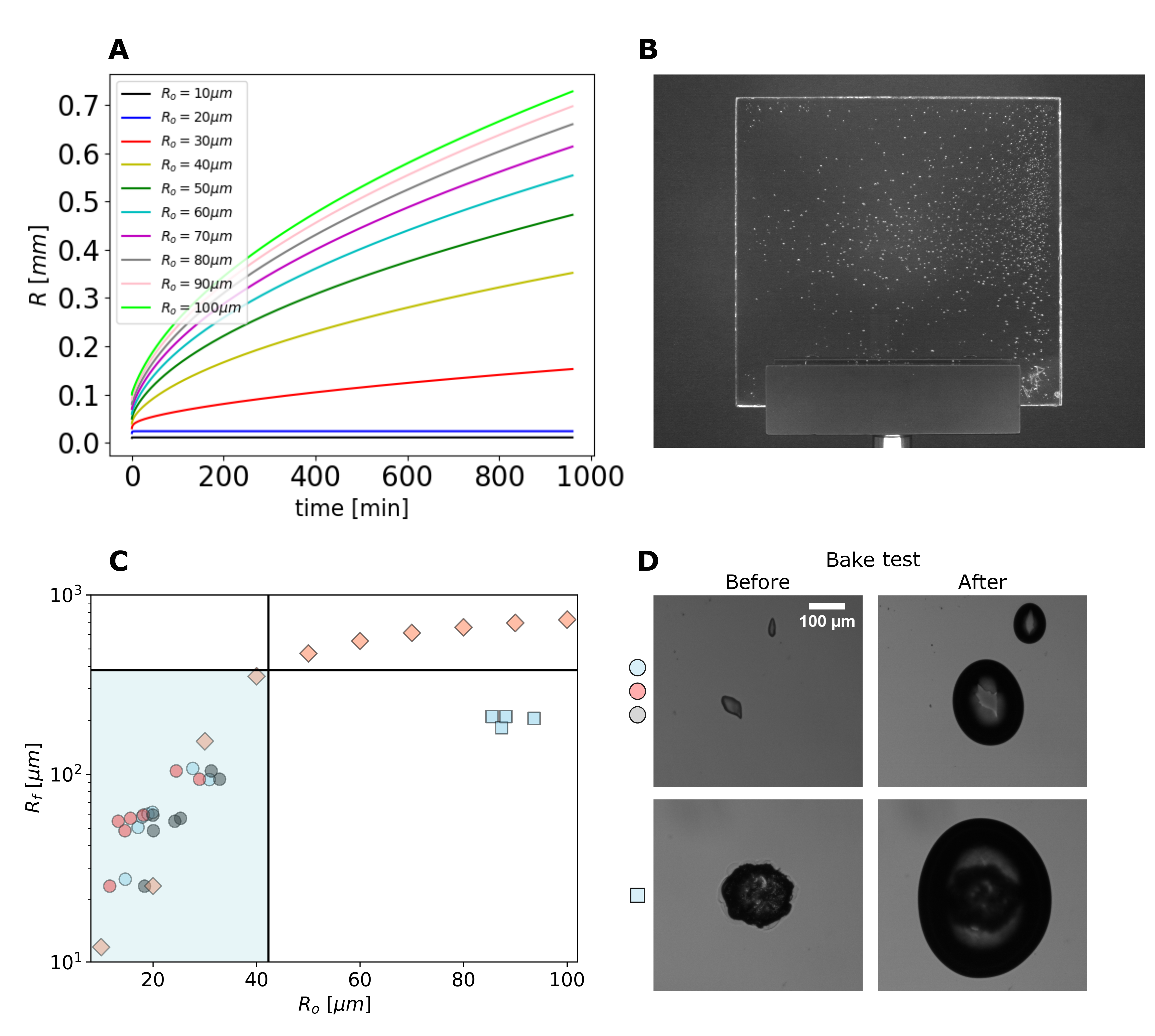}
\centering
\caption{When \textcolor{blue}{water} is \textit{partly} to blame for bubbles in LSG during bake testing (RB41, 16hrs, 100°C). Bubbles may form if the LSG sample is populated by sufficiently large nuclei despite ideal de-airing conditions. \textbf{A} Simulation results showing the aforementioned \textit{size effect}, i.e., only nuclei above a critical size can grow into a visible bubble. \textbf{B} Photograph of sample of LSG (10 x 10 cm) post bake testing. \textbf{C} Comparison between model (\textcolor{Apricot}{\scalebox{0.8}{$\blacklozenge$}}) and measurements from natural (\textcolor{RedOrange}{\scalebox{1.2}{$\bullet$}}\textcolor{SkyBlue}{\scalebox{1.2}{$\bullet$}}\textcolor{gray}{\scalebox{1.2}{$\bullet$}}) and artificial inclusions (\textcolor{SkyBlue}{\scalebox{0.7}{$\blacksquare$}}). The different marker colors represent the same natural gaseous inclusions, albeit using different approximations to determine $R_0, R_f$. The shaded area delineates the region where our bubble model is expected to be quantitatively effective. \textbf{D} Examples of microscope images of gaseous inclusions used to create figure \textbf{C} with the corresponding markers pertaining to their origin, i.e., either natural or artificial.}
\label{fig7}
\end{figure*}

\hyperref[fig7]{Figure~7}A shows the results of the bake test simulation. Very small nuclei, below $30\mu$m in size, cannot significantly expand during the bake test and stay below the lower limit of resolution of the human eye. A sample of LSG populated with such nuclei will thus surely always pass the test. Nuclei of intermediate sizes, of around $40-60\mu$m, can reach sizes between 0.3-0.5mm and will therefore, depending on the observer, result in possible test failure. Large nuclei ($>70\mu$m), however, grow to sizes between 0.5-0.7mm and will most definitely result in test failure regardless of the observer. 

As qualitative confirmation of the aforementioned results, \hyperref[fig7]{figure~7}B presents a picture of one of our many bubbling samples of LSG post bake testing. The final size of the bubbles that grew therein agree with the order of magnitude of the simulation, i.e., $\sim 1$mm.

To better gauge the efficacy of our model, \hyperref[fig7]{figure~7}C compares the predicted final bubble size $R_f$\textemdash as a function of the initial gaseous inclusion size $R_0$\textemdash against their experimental counterparts. For the latter, we selected gaseous inclusions having a morphology similar to those portrayed in \hyperref[fig5]{figure~5}D, i.e., partially dissolved non-spherical bubbles (\hyperref[fig7]{figure~7}D [top row]). We additionally included measurements from artificial inclusions trapped in cylindrical crenels that were etched on the glass surface (\hyperref[fig7]{figure~7}D [bottom row]).  When determining $R_0$, $R_f$, two approximations were made.

\textbf{2-D approximation}. The observed projection of the non-spherical inclusion/bubbles was approximated as an ellipse. We manually determined the semi-major axes $a_{0,f}$, $b_{0,f}$, wherein $a_{0,f}>b_{0,f}$, and the indices $0,f$ denote, as before, conditions before and after bake testing respectively. The value of $R_{0,f}$ corresponds to a circle of equivalent area as to that of the measured ellipse in 2-D. We thus have $R_{0,f}=\sqrt{a_{0,f}\times b_{0,f}}$ (blue circles in \hyperref[fig7]{figure~7}C). The same approximation was used for the artificial inclusions (blue squares in \hyperref[fig7]{figure~7}C).

\textbf{3-D approximation.} We approximate the gaseous inclusions/bubbles as ellipsoid caps with a corresponding additional semi-major axis $c_{0,f}$. $R_{0,f}$ is then the radius of a spherical bubble in 3-D space of equivalent volume as that of the ellipsoid cap. We thus have $R_{0,f}=\sqrt[3]{(a_{0,f} \times b_{0,f}\times c_{0,f})}$. Meanwhile, $c_{0,f}$ was estimated as follows. Post bake testing, the gaseous inclusions/bubbles evolve from being highly ellipsoidal with $a_0/b_0\approx 3.1$ to becoming circular like since $a_f/b_f\approx 1.3$. Because both $a_f,b_f\ll h$ (PVB half thickness, 380$\mu$m), we can rightly assume that the resulting 2-D symmetry between the semi-major axes $a_f, b_f$ extends to $c_f$ as well. Conservatively, we thusly take $c_f\approx b_f$ when computing $R_f$. Prior to bake testing, we find that $a_0 \approx 35 \mu$m is in the same order of magnitude as $e$ (PVB roughness, $\sim40 \mu$m), whereas $b_0$ is roughly one third of this value. Heuristically, we assumed symmetry with either semi-major axis to provide bounds for $R_0$. The upper bound of the ellipsoid volume\textemdash and evidently $R_0$\textemdash is set by $c_o\approx a_0$ (grey circles in \hyperref[fig7]{figure~7}C). This bound tacitly implies the polymer roughness was relatively unaffected during autoclaving for the gaseous inclusion in question. Meanwhile, a lower bound for $R_0$ is given by $ c_o\approx b_0$ (red circles in \hyperref[fig7]{figure~7}C).

Overall, as shown in \hyperref[fig7]{figure~7}C, the hereinabove described approximations for $R_o, R_f$ cluster around similar values but most importantly, align closely with the trend set by the bubble model (orange markers in \hyperref[fig7]{figure~7}C). In particular, this holds in the region where we expect our model to be most quantitatively valid. This is represented as a shaded region in the figure, which is bounded by a nuclei of size $R_0\approx 42 \mu$m for which the final bubble size is predicted to match the PVB thickness, namely, $R_f=h$. Beyond this region, we expect confinement effects to play a major role on bubble kinetics, and thus, on the final bubble size. Indeed, for larger nuclei and as demonstrated by the artificial inclusions (blue squares in \hyperref[fig7]{figure~7}C), the model overshoots the experimental data. The overall experimental trend, nevertheless, matches the simulated one. 

Therefore, we reach a first conclusion regarding bake test failure: even if invisible to the human eye, the presence of large nuclei post lamination in a sample of LSG can result in (\textit{partly}) water-driven bubble growth (the bubbles in the simulation can reach $\color{blue}{x_w}$~$\sim0.18$). This holds for excellent de-airing and regardless of whether the PVB polymer is meticulously prepared at standard humidity conditions. To reiterate, failure in this first case is driven by a combination of heterogeneous nucleation, critical nuclei size, and water working in tandem with air to form visible bubbles. 

\section{When Air is the \textit{Main} Culprit}\label{sec:air_main}

\begin{figure*}[htp]
\includegraphics[width=0.84\textwidth]{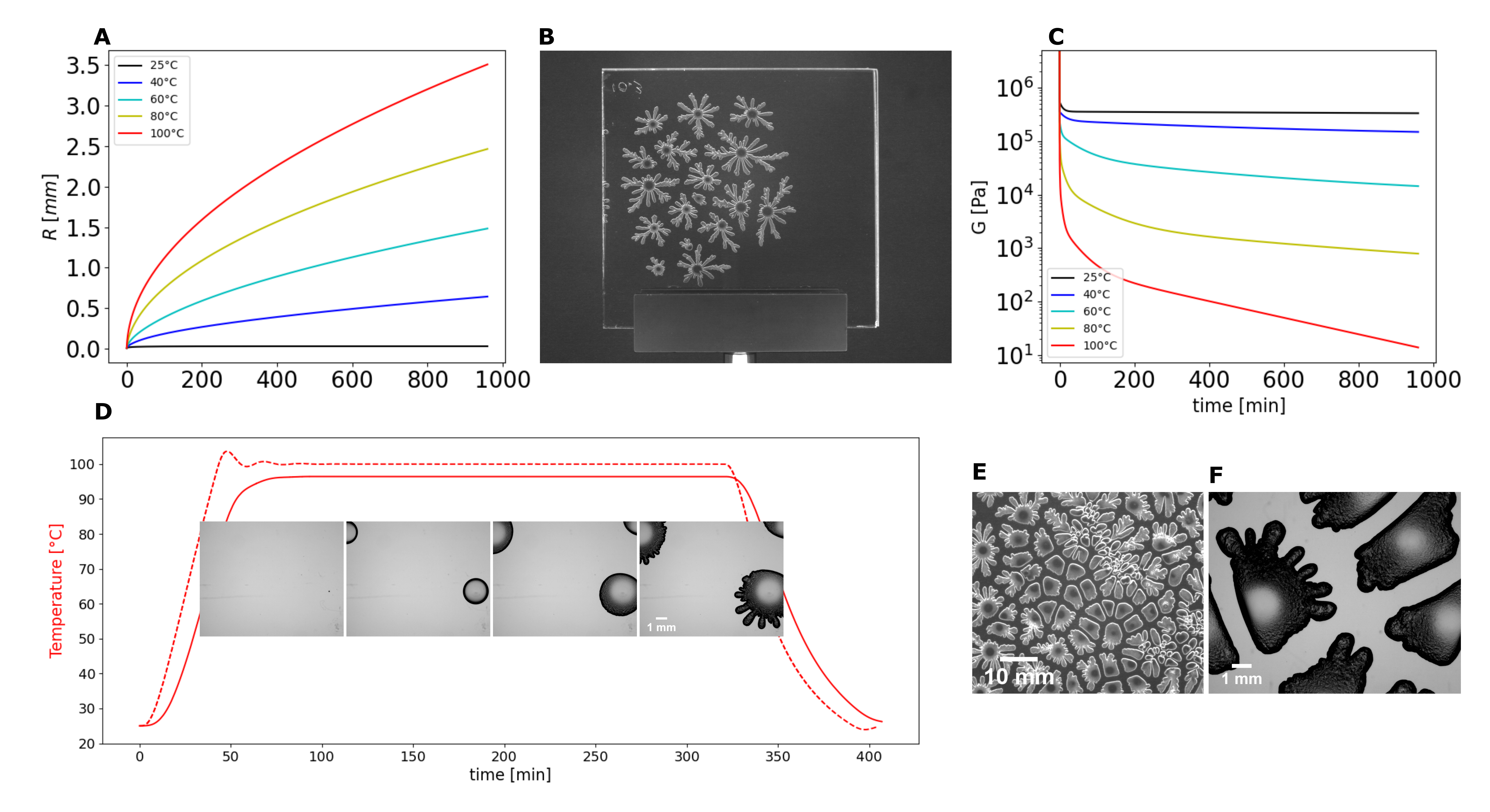}
\centering
\caption{When \textcolor{red}{air} is the \textit{main} culprit behind bubbles during bake testing because of anomalous oversaturation (16hrs, 100°C). \textbf{A} Simulation results for various temperatures. \textbf{B} Photograph of a LSG sample (RB11, 10 x 10 cm) after bake testing showing a type of window frost instability as the bubble size greatly exceeds the polymer thickness under confinement and high adhesion. \textbf{C} PVB relaxation as a function of temperature.
\textbf{D} Short bake test using the autoclave set-up. \textcolor{red}{\rule[.5ex]{0.5em}{.5pt}}\  \textcolor{red}{\rule[.5ex]{0.5em}{.5pt}}, \textcolor{red}{\rule[.5ex]{1em}{.5pt}} Measured temperature towards the glass edge and center respectively. \textbf{E},\textbf{F} Catastrophic window-frost-like failure with RB11. Bubbles do not coalesce.} 
\label{fig8}
\end{figure*}

During lamination, the amount of air dissolved in the PVB bulk can exceed that which is originally present in the interfacial bubbles from the pre-press. We refer to this condition as \textit{anomalous air oversaturation}. The reasons are manifold and beyond the scope of the present work, but the condition is extremely detrimental\textemdash bubble wise\textemdash in LSG. 

We induced anomalous air oversaturation in glass sandwiches (30 x 50 cm) of RB11 (thickness 0.38 mm, $a=0.25$) where we prematurely sealed three edges of the sample by heat and pressure. By subsequent calendering along the longest edge (50 cm), we trapped high quantities of excess air in a region equal to 1/5 of the total length. Such samples were autoclaved and the trap region subdivided into three 10 x 10 cm samples for bake testing. The driver in this case is $\textcolor{red}{ f_a }\sim10$ at room conditions. 

When performing the same numerical sweep in terms of initial bubble size as in the previous section, we found that bubbles grow irrespective of initial size during the bake test (not shown). The bubbles are also comprised mostly by air ($\color{red}x_a$ ~$\sim0.92$). For clarity, in \hyperref[fig8]{figure~8}A we present only the curve for the smallest nucleus $R_o=10\mu$m (red line). Compared to the (worse) water-driven case at $R_o=100\mu$m, bubbles growing under anomalous air conditions can reach large sizes in a short period of time. In the example at hand, the difference in size is factor of five, and the timescale is in the order of mins compared to hours. This suggests that bubble size (and onset of visibility) is a proxy indicator for the gas responsible behind bubble growth.  Finally, notice that the predicted bubble size for anomalous air oversaturation is much larger than the PVB thickness $R/h\sim 18$! At these sizes, our physical description is quantitatively questionable since the assumption of spherical symmetry is violated. Indeed as shown in \hyperref[fig8]{figure~8}B, bubbles do not maintain sphericity but rather a beautiful instability, similar to window frost or snowflakes, is triggered instead. Nonetheless, the general truths obtained from the model are unchanged: in anomalous air oversaturation conditions, bubbles can reach large sizes in a short amount of time during bake testing at 100°C.   

Because anomalous air oversaturation can be so catastrophic for LSG, we investigated in \hyperref[fig8]{figure~8}A the effect of different temperature levels on bubble growth, seeing that the nucleus size was of less importance. We tested the already discussed temperature of 100°C, down to 80, 60, 40, and 25°C all with a timescale of 16hrs. At room conditions, the nucleus is stable and there is no indication from the figure that significant bubble growth may occur (black line). In this case, the high stiffness of the polymer (\hyperref[fig8]{figure~8}C) forcibly stabilizes the nucleus and prevents it from growing as expected from eq.~\ref{eq:bubble_radius_model}. However, the moment the temperature is increased, to say 40°C (blue line), bubbles can reach a visible size ($0.5mm$) in as little as 16hrs. Evidently, and as predicted as well by eq.~\ref{eq:bubble_radius_model}, the higher the temperature, the bigger the bubble. 

\hyperref[fig7]{Figure~8}D presents the results of a short bake test (4 hrs) using our transparent autoclave where indeed, and in qualitative agreement with our model, we observed bubble growth after barely $\sim$42 mins or $T\sim75$°C which is much sooner than the target temperature of the test (95-100°C). The bubble initially expands in circular fashion, until a size of roughly $R\sim 1.9mm \ (R/h\sim 10)$, after which the instability kicks in. The bubble develops branching arms which is a strong indication that our model, despite being qualitatively adequate, is not quantitatively applicable in this situation. First and foremost, the high level of deformation, even before the instability appears, implies the relaxation modulus cannot be presumed as independent of the applied strain. Regrettably, relaxation data for PVB under large deformation\textemdash for elevated temperatures and long timescales\textemdash is unavailable in the existent literature; an unfortunate situation that renders our model for bubbles driven by anomalous air oversaturation in LSG as qualitative only.

\textbf{On the nature of the instability}. Mechanical instabilities in polymers may stem from a plurality of reasons. Often, they are triggered whenever the rate of deformation is faster than the polymer relaxation time~\cite{tabuteau2009microscopic}, are influenced by the initial size of the nucleus/imperfection~\cite{kundu2009cavitation}, may be impacted by the aspect ratio of the polymer sheet~\cite{guo2023crack}, or even the presence of neighboring bubbles~\cite{haudin2016bubble}. 

In our case, the instability is seemingly and primarily a mix of confinement and high adhesion effects. As the gas expands, it burrows inside the polymer and triggers the instability because delamination never occurs. \citet{saintyves2013bulk} have studied this type of instability\textemdash in connection with soft matter materials\textemdash where they have indeed identified the drivers behind it: confinement, adhesion, and a critical strain level. Secondarily, the instability is influenced by a property of bubbles in safety glass which we discussed in passing in $\S$~\ref{sec:toy_exp}. Notice in \hyperref[fig8]{Figure~8}E how the irregular bubble structures are densely packed, yet they never coalesce because the PVB polymer never delaminates from the glass. Each bubble, despite its inherent anisotropy, maintains a distinct shape that is strongly affected by the presence of neighboring bubbles (see \hyperref[fig8]{figure~8}F).  

\section{Discussion and Concluding Remarks}\label{sec:conclusion}

Our results indicate that a finished sample of laminated safety glass is in a metastable state from the point of view of bubble formation. The PVB bulk is oversaturated with air post lamination, meaning that bubbles are thermodynamically favorable to being with in these assemblies, i.e., their natural tendency is indeed to bubble. It is only the stiff nature of the polymer at room conditions that prevents them from doing so. How long before (and whether) they will bubble at all depends on the nature of the polymer itself (e.g., composition, thickness), the level of air oversaturation, the temperatures faced by the glass assembly, and the presence (or lack thereof) of gaseous nuclei and their size. 

We discussed two cases for bubble growth during bake testing, when \textcolor{blue}{water} is \textit{partly} to blame, and when \textcolor{red}{air} is the \textit{main} culprit. A passing juxtaposition of \hyperref[fig7]{figures~7}A and \hyperref[fig8]{8}A reveals some important distinctions between the two cases. The expected bubble size is markedly different and low-temperature bubbles are more likely in one case than in the other. These bubble features may qualitatively hint at the gas most responsible for bubble formation in industrial settings. In short, large bubbles that appear rapidly, or a cold temperature, are most likely stemming from air-related problems, rather than water-related issues. The current test, in this regard, does not discriminate between these two mechanisms.

On purpose, we did not discuss a third scenario for bubble formation during bake testing: when water is the \textit{main} culprit. In this situation, the PVB polymer contains a large amount of water dissolved in the bulk compared to standard conditions. The omission of this case responds to a technical reason. It is a relatively simple matter to non-destructively measure the amount of water in the PVB bulk post lamination (see figure S3 in the suppl. mat.), and thus, appraise if water is \textit{fully} to blame.

Setting aside the above discussion, we have presented a physical description for bubble growth that deals with multiple gases, thermal effects, as well as a complex time-temperature influenced rheology. The description highlights the cross-disciplinary nature of bubble phenomena and can be extended to other polymer systems as well. 

Logically, our physical model is directly applicable to Maxwell-type polymers in general, such as molten PDMS or polyisoprene, which display\textemdash like PVB\textemdash time-temperature dependent vanishing shear relaxations. Viscoelastic solids are equally concerned. The latter are adeptly described via the generalized Maxwell model by the addition of a single spring constant $G_\infty$ that reflects the asymptotic, elastic behavior of the polymer. This means that the bubbling behavior of crosslinked polymers, of which ethyl vinyl acetate (EVA) is a great example, can equally be modeled under our framework.  Finally, our bubble model is equally applicable to more classical polymers, such as elastomers (e.g., natural rubber) or hydrogels since the latter can be treated as viscoelastic solids with an infinite relaxation time. 

\section*{Author contributions}
CAM, KP, and EL conceived the research project and drafted/edited the research paper. CAM carried out the experimental campaigns, performed the necessary analysis, developed the physical model, and associated python codes. CAM/EL devised the experimental set-up. KP conceived the experimental settings for anomalous air oversaturation. CAM/KP worked jointly in the setup of the simulations.

\section*{Conflicts of interest}
There are no conflicts to declare.

\section*{Data availability}

The data supporting this article have been included as part of the Supplementary Information. All other remaining data is available upon reasonable request to the corresponding author.

\section*{Acknowledgements}
We gratefully thank Jerome Giraud for helping design and build the experimental autoclave set-up, Lionel Bureau for facilitating the equipment for bake testing, and Bruno Travers for his aid in the instrumentation of temperature sensors. We also extend our deepest gratitude to Vincent Rachet and Nicolas Nadaud for their interest and support during our study, as well as Richard Villey and Anne-Laure Vayssade for fruitful discussions around bubbles in glass.

\balance

\bibliography{bib} 

@article{epstein1950stability,
  title={On the stability of gas bubbles in liquid-gas solutions},
  author={Epstein, Paul S and Plesset, Milton S},
  journal={The Journal of Chemical Physics},
  volume={18},
  number={11},
  pages={1505--1509},
  year={1950},
  publisher={American Institute of Physics}
}

@article{nascimento2016first,
  title={On the first patents, key inventions and research manuscripts about glass science \& technology},
  author={Nascimento, Marcio Luis Ferreira and Zanotto, Edgar Dutra},
  journal={World Patent Information},
  volume={47},
  pages={54--66},
  year={2016},
  publisher={Elsevier}
}

@book{panati2016panati,
  title={Panati's extraordinary origins of everyday things},
  author={Panati, Charles},
  year={2016},
  publisher={Chartwell Books}
}

@article{carrot2016polyvinyl,
  title={Polyvinyl butyral},
  author={Carrot, Christian and Bendaoud, Amine and Pillon, Caroline and Olabisi, O and Adewale, K},
  journal={Handbook of thermoplastics},
  volume={2},
  pages={89--137},
  year={2016},
  publisher={CRC press}
}

@article{liger2008recent,
  title={Recent advances in the science of champagne bubbles},
  author={Liger-Belair, G{\'e}rard and Polidori, Guillaume and Jeandet, Philippe},
  journal={Chemical Society Reviews},
  volume={37},
  number={11},
  pages={2490--2511},
  year={2008},
  publisher={Royal Society of Chemistry}
}

@article{webb2011effect,
  title={The effect of temperature and viscoelasticity on cavitation dynamics during ultrasonic ablation},
  author={Webb, Ian R and Payne, Stephen J and Coussios, Constantin-C},
  journal={The Journal of the Acoustical Society of America},
  volume={130},
  number={5},
  pages={3458--3466},
  year={2011},
  publisher={AIP Publishing}
}

@article{yang2005model,
  title={A model for the dynamics of gas bubbles in soft tissue},
  author={Yang, Xinmai and Church, Charles C},
  journal={The Journal of the Acoustical Society of America},
  volume={118},
  number={6},
  pages={3595--3606},
  year={2005},
  publisher={AIP Publishing}
}

@article{fyrillas2000factors,
  title={Factors determining the stability of a gas cell in an elastic medium},
  author={Fyrillas, MM and Kloek, W and Van Vliet, T and Mellema, J},
  journal={Langmuir},
  volume={16},
  number={3},
  pages={1014--1019},
  year={2000},
  publisher={ACS Publications}
}

@article{kloek2001effect,
  title={Effect of bulk and interfacial rheological properties on bubble dissolution},
  author={Kloek, William and van Vliet, Ton and Meinders, Marcel},
  journal={Journal of Colloid and interface Science},
  volume={237},
  number={2},
  pages={158--166},
  year={2001},
  publisher={Elsevier}
}

@article{saintyves2013bulk,
  title={Bulk elastic fingering instability in hele-shaw cells},
  author={Saintyves, Baudouin and Dauchot, Olivier and Bouchaud, Elisabeth},
  journal={Physical review letters},
  volume={111},
  number={4},
  pages={047801},
  year={2013},
  publisher={APS}
}

@article{arauz2023water,
  title={Water Clustering in Polyvinyl Butyral (PVB): Evidenced by Diffusion and Sorption Experiments},
  author={Arauz-Moreno, C and Piroird, K and Lorenceau, E},
  journal={The Journal of Physical Chemistry B},
  volume={127},
  number={51},
  pages={11064--11073},
  year={2023},
  publisher={ACS Publications}
}

@article{arauz2022extended,
  title={Extended time--temperature rheology of polyvinyl butyral (PVB)},
  author={Arauz Moreno, Carlos and Piroird, Keyvan and Lorenceau, Elise},
  journal={Rheologica Acta},
  volume={61},
  number={8},
  pages={539--547},
  year={2022},
  publisher={Springer}
}

@article{arauz2025champagne,
  title={From champagne to confined polymer: Natural and artificial bubble nucleation},
  author={Arauz-Moreno, Carlos and Piroird, Keyvan and Lorenceau, Elise},
  journal={Physical Review E},
  volume={112},
  number={1},
  pages={015424},
  year={2025},
  publisher={APS}
}

@article{lopez2019mechanical,
  title={Mechanical characterization of polyvinil butyral from static and modal tests on laminated glass beams},
  author={L{\'o}pez-Aenlle, M and Noriega, A and Pelayo, F},
  journal={Composites Part B: Engineering},
  volume={169},
  pages={9--18},
  year={2019},
  publisher={Elsevier}
}

@article{centelles2021viscoelastic,
  title={Viscoelastic characterization of seven laminated glass interlayer materials from static tests},
  author={Centelles, Xavier and Pelayo, Fern{\'a}ndez and Lamela-Rey, Mar{\'\i}a Jes{\'u}s and Fern{\'a}ndez, A In{\'e}s and Salgado-Pizarro, Rebeca and Castro, J Ramon and Cabeza, Luisa F},
  journal={Construction and Building Materials},
  volume={279},
  pages={122503},
  year={2021},
  publisher={Elsevier}
}

@article{yang2024deformation,
  title={Deformation and Fracture Behaviors of Tough Plasticized Poly (vinyl butyral) across Broad Temperature and Strain Rate Ranges},
  author={Yang, Erjie and Ji, Rongyao and Miao, Jibin and Yan, Qi and Li, Liangbin and Guo, Hang and Cui, Kunpeng},
  journal={Macromolecules},
  volume={57},
  number={16},
  pages={8123--8133},
  year={2024},
  publisher={ACS Publications}
}

@article{pauli2024simplified,
  title={Simplified approach for modeling standard PVB at large deformations and long-term loading},
  author={Pauli, Alexander and Siebert, Geralt},
  journal={Glass Structures \& Engineering},
  volume={9},
  number={1},
  pages={59--73},
  year={2024},
  publisher={Springer}
}

@article{corroyer2013characterization,
  title={Characterization of commercial polyvinylbutyrals},
  author={Corroyer, Elsa and Brochier-Salon, Marie-Christine and Chaussy, Didier and Wery, S{\'e}bastien and Belgacem, Mohamed Naceur},
  journal={International Journal of Polymer Analysis and Characterization},
  volume={18},
  number={5},
  pages={346--357},
  year={2013},
  publisher={Taylor \& Francis}
}

@article{chen2022pummel,
  title={On the pummel test and pummel-value evaluation of polyvinyl butyral laminated safety glass},
  author={Chen, Xing and Schuster, Miriam and Schneider, Jens},
  journal={Composite Structures},
  volume={280},
  pages={114878},
  year={2022},
  publisher={Elsevier}
}

@article{del2016determining,
  title={Determining material response for Polyvinyl Butyral (PVB) in blast loading situations},
  author={Del Linz, Paolo and Wang, Yi and Hooper, Paul A and Arora, Hari and Smith, David and Pascoe, Luke and Cormie, David and Blackman, Bamber RK and Dear, John P},
  journal={Experimental Mechanics},
  volume={56},
  pages={1501--1517},
  year={2016},
  publisher={Springer}
}

@incollection{belis2019architectural,
  title={Architectural glass},
  author={Belis, Jan and Louter, Christian and Nielsen, Jens H and Schneider, Jens},
  booktitle={Springer handbook of glass},
  pages={1781--1819},
  year={2019},
  publisher={Springer}
}

@inproceedings{stevels2020determination,
  title={Determination and verification of PVB interlayer modulus properties},
  author={Stevels, Wim and D'Haene, Pol},
  booktitle={Challenging Glass Conference Proceedings},
  volume={7},
  year={2020}
}

@article{samieian2019bonding,
  title={On the bonding between glass and PVB in laminated glass},
  author={Samieian, Mohammad Amin and Cormie, David and Smith, David and Wholey, Will and Blackman, Bamber RK and Dear, John P and Hooper, Paul A},
  journal={Engineering Fracture Mechanics},
  volume={214},
  pages={504--519},
  year={2019},
  publisher={Elsevier}
}

@article{fourton2020adhesion,
  title={Adhesion rupture in laminated glass: influence of adhesion on the energy dissipation mechanisms},
  author={Fourton, Paul and Piroird, Keyvan and Ciccotti, Matteo and Barthel, Etienne},
  journal={Glass Structures \& Engineering},
  volume={5},
  pages={397--410},
  year={2020},
  publisher={Springer}
}

@article{chen2023effect,
  title={Effect of moisture on the delamination properties of fractured PVB-laminated glass: A joint experimental and numerical study},
  author={Chen, Xing and Lin, Binbin and Schuster, Miriam and Chen, Suwen and Xu, Bai-Xiang and Schneider, Jens},
  journal={Composite Structures},
  volume={322},
  pages={117381},
  year={2023},
  publisher={Elsevier}
}

@article{botz2019experimental,
  title={Experimental investigations on the creep behaviour of PVB under different temperatures and humidity conditions},
  author={Botz, M and Wilhelm, K and Siebert, G},
  journal={Glass Structures \& Engineering},
  volume={4},
  number={3},
  pages={389--402},
  year={2019},
  publisher={Springer}
}

@article{desloir2019plasticization,
  title={Plasticization of poly (vinyl butyral) by water: Glass transition temperature and mechanical properties},
  author={Desloir, Marl{\`e}ne and Benoit, Cyril and Bendaoud, Amine and Alcouffe, Pierre and Carrot, Christian},
  journal={Journal of Applied Polymer Science},
  volume={136},
  number={12},
  pages={47230},
  year={2019},
  publisher={Wiley Online Library}
}

@inproceedings{stevels2024ten,
  title={Ten Years of Stiff PVB: An Overview of Developments and Current Status},
  author={Stevels, Wim},
  booktitle={Challenging Glass Conference Proceedings},
  volume={9},
  year={2024}
}

@article{elziere2019supramolecular,
  title={Supramolecular structure for large strain dissipation and outstanding impact resistance in polyvinylbutyral},
  author={Elziere, Paul and Fourton, Paul and Demassieux, Quentin and Chenneviere, Alexis and Dalle-Ferrier, C{\'e}cile and Creton, Costantino and Ciccotti, Matteo and Barthel, Etienne},
  journal={Macromolecules},
  volume={52},
  number={20},
  pages={7821--7830},
  year={2019},
  publisher={ACS Publications}
}

@inproceedings{muller2024imperfections,
  title={Imperfections in Laminated Safety Glass: An Experimental Case Study},
  author={M{\"u}ller, Paul and Schuler, Christian and Gr{\"o}tzner, Jakob and Dix, Steffen and Hiss, Stefan},
  booktitle={Challenging Glass Conference Proceedings},
  volume={9},
  year={2024}
}

@inproceedings{zhou2024post,
  title={Post-Failure Behavior of Point-Fixing Laminated Glass Plates under Out-of-Plane Uniform Pressure},
  author={Zhou, Sicheng and Cattaneo, Sara and Biolzi, Luigi},
  booktitle={Challenging Glass Conference Proceedings},
  volume={9},
  year={2024}
}

@article{haraya1992permeation,
  title={Permeation of oxygen, argon and nitrogen through polymer membranes},
  author={Haraya, Kenji and Hwang, Sun-Tak},
  journal={Journal of membrane science},
  volume={71},
  number={1-2},
  pages={13--27},
  year={1992},
  publisher={Elsevier}
}

@article{tabuteau2009microscopic,
  title={Microscopic mechanisms of the brittleness of viscoelastic fluids},
  author={Tabuteau, Herv{\'e} and Mora, Serge and Porte, Gr{\'e}goire and Abkarian, Manouk and Ligoure, Christian},
  journal={Physical review letters},
  volume={102},
  number={15},
  pages={155501},
  year={2009},
  publisher={APS}
}

@article{kundu2009cavitation,
  title={Cavitation and fracture behavior of polyacrylamide hydrogels},
  author={Kundu, Santanu and Crosby, Alfred J},
  journal={Soft Matter},
  volume={5},
  number={20},
  pages={3963--3968},
  year={2009},
  publisher={Royal Society of Chemistry}
}

@article{guo2023crack,
  title={On crack nucleation and propagation in elastomers: I. In situ optical and X-ray experimental observations},
  author={Guo, Jinlong and Ravi-Chandar, Krishnaswamy},
  journal={International Journal of Fracture},
  volume={243},
  number={1},
  pages={1--29},
  year={2023},
  publisher={Springer}
}

@article{haudin2016bubble,
  title={Bubble dynamics inside an outgassing hydrogel confined in a Hele-Shaw cell},
  author={Haudin, Florence and Noblin, Xavier and Bouret, Yann and Argentina, M{\'e}d{\'e}ric and Raufaste, Christophe},
  journal={Physical Review E},
  volume={94},
  number={2},
  pages={023109},
  year={2016},
  publisher={APS}
}
\bibliographystyle{rsc}

\end{document}